# A GROUPED SYSTEM ARCHITECTURE FOR SMART GRIDS BASED AMI COMMUNICATIONS OVER LTE

Mahmoud M. Elmesalawy[1] and A.S.Ali[1]

[1]Department of Electronics, Communications, and Computer Engineering, Helwan University, Cairo, Egypt

## ABSTRACT

*A smart grid based Advanced Metering Infrastructure (AMI), is a technology that enables the utilities to monitor and control the electricity consumption through a set of various smart meters (SMs) connected via a two way communication infrastructure. One of the key challenges for smart grids is how to connect a large number of devices. On the other hand, 4G Long Term Evolution (LTE), the latest standard for mobile communications, was developed to provide stable service performance and higher data rates for a large number of mobile users. Therefore, LTE is considered a promising solution for wide area connectivity for SMs. In this paper, a grouped hierarchal architecture for SMs communications over LTE is introduced. Then, an efficient grouped scheduling technique is proposed for SMs transmissions over LTE. The proposed architecture efficiently solves the overload problem due to AMI traffic and guarantees a full monitoring and control for energy consumption. The results of our suggested solution showed that LTE can serve better for smart grids based AMI with particular grouping and scheduling scheme. In addition, the presented technique can able to be used in urban areas having high density of SMs.*

## 1. INTRODUCTION

Nowadays, it is highly necessity to enable the electric utilities in all countries to fully monitor and control customer's consumptions. Therefore, data collection, communication and management system with high temporal efficiency must be applied in order to optimize the energy efficiency. To realize this objective, the current power grid must be evolved to be smart one. Smart grid integrates the Information and Communication Technology (ICT) into the existing power grid to convert it to an intelligent power grid. As a result, it can monitor the grid status, power transmission, and customer's consumptions through a system of sensors, communication network and software applications [1]. Consequently, smart grid is mainly depending on communication in order to coordinate the generation, distribution, and consumption of customer electricity usage [2]. Fully monitored and controlled power grid lead to enhance energy efficiency of the grid, adjusting the power consumption of household applications to save energy, control load drop in peak time, and reduce energy losses.

Advanced Metering Infrastructure, one of the key components in smart grid, consists of Automatic Meter Reading (AMR), a set of actuator and a two-way communication system. Based on its capability of instantaneous data transmission and imposing consumptions of customers, AMI enables smart grid to manage consumption demands [1]. The data message of AMI contains the measurements of the consumption from customers including, meter reading, device ID, time stamps and other identification information about the AMI and the customer [3]. Several research works in literature define the various measurements and communication requirements for different smart grid applications in terms of delay, reliability, bandwidth, and more generally quality of service (QoS) attributes [4], [5], [6]. The different communication technologies and infrastructures used for smart grids are deliberated in [7-11]. As regards the data rate of AMI, each consumption measurement is sent every 10 ~ 15 minutes, therefore AMI needs to report consumption status at a rate of 4 ~ 6 times per hour [3].

A number of key challenges are featured in smart grid to achieve the role required from it. One of these challenges is the real time data gathering, data transmission, and data processing which lead to real time trace and control. Another challenge is how to connect a large number of SMs, controllers, actuators and data collectors, concentrators and storages. Therefore, a high efficient and reliable communication network is needed to achieve the required real time monitor and control system for smart grids. Different wireless communication technologies used for smart grids are presented in [12-16]. Current cellular mobile networks can be employed to carry the large simultaneous data collected from AMI end points (smart meters) but with congestion and competition at the Radio Access Networks (RANs) [17]. This congestion will deteriorate the mobile network performance in form of increasing number of packet loss, unbearable delay which greatly affects the required services and QoS for the mobile users in the network.

4G Long Term Evolution (LTE), defined by 3rd Generation Partnership Project (3GPP) is the latest cellular mobile network technology. It is promising to support mobile users' demands with stable service performance and high data rates (300 Mbps in the downlink, and 75 Mbps in the uplink) [18]. LTE was developed as a standard for mobile communications such as voice, video, web browsing, and other applications which have different traffic type than those of the smart grid. Therefore, it seems that LTE networks are not suitable for smart grid data communications because it has not been designed for smart grid applications. Actually there are no Resource Blocks (RBs) are reserved for smart grid data at the evolved NodeB (eNodeB). However, LTE has become a promising candidate for the smart grid data communication network. Several LTE scheduler techniques for mobile User Equipments (UEs) are introduced in literatures [19-22]. Once more, these techniques do not take into account the smart grid traffic which has significant different requirements compared to the services developed to mobile users. Consequently, and according to the best of our knowledge, there is no a standard LTE scheduling technique designed to handle the traffic for mobile users services hand in hand with smart grid traffic maintaining the required QoS for these users services.

In this paper, we introduce a grouped hierarchical architecture for SMs communications in smart grid over LTE infrastructure. Then a grouped scheduling technique in LTE supporting AMI communications for smart grids is presented. The key idea of our suggested solution is to enable LTE network to support a real time data transmission of the smart grid even in case of heavily loaded LTE network and high density smart grid areas with a huge number of SMs to be served. The rest of this paper is organized as follows. Section 2, introduces the suggested grouped network architecture design. In section 3, the system model is presented. The results and discussion is provided in section 4. Section 5, concludes the paper.

## 2. THE SUGGESTED GROUPED SYSTEM ARCHITECTURE

In this section, the overall architecture of the proposed solution is discussed. Then a brief comparison with the flat architecture will be presented to show the advantages of our technique over the flat scenario. The flat architecture in which the SMs in customers' site are directly connected to the LTE network is depicted in Figure 1. In this scenario, an LTE communication module is used at each SM to communicate directly to the eNodeB covered his area. Actually the geographical area is divided into cells, each cell under control of eNodeB. Each eNodeB provides the radio communication resources to SMs according to the available RBs. Recalling that the eNodeB does not have the powers in terms of reserved RBs to serve that such large number of SMs, therefore there is an urgent need to find a solution to this growing problem. We demonstrate that our suggested solution, grouped system architecture and scheduling technique, will support in solving this problem.

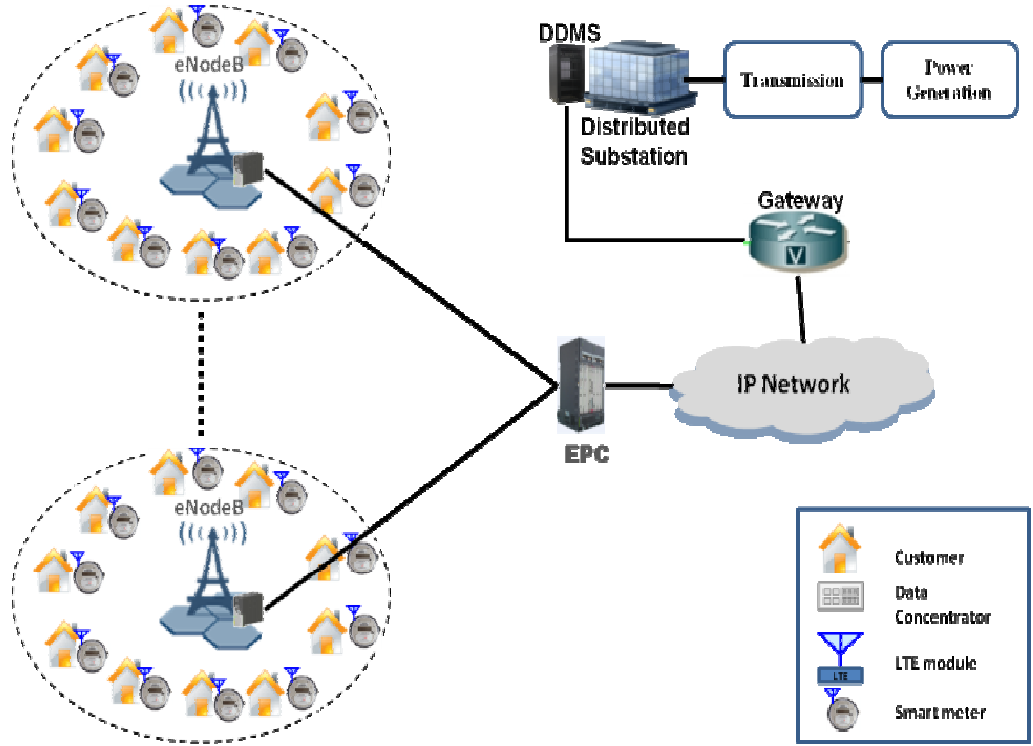

Figure 1. The flat network architecture.

The suggested grouped system architecture is depicted in Figure 2. In this grouped architecture the total number of smart meters ($K$) which must be served with a single eNodeB will be divided into "$N$" groups. The SMs in each group are connected to its Data Concentrator (DC) through a local communication network. The DC, a key element in our solution, includes Data Collector and Management System (DCMS), and two types of communication networks; one for communication with the SMs in his group (Zigbee, as an example) and the other for communication with eNodeB in LTE network. Each DC, through its local network, will receive periodically the AMI messages from each SM in his group. These messages will be concentrated in one message, Total Consumption (TC message), which has the total consumption of all customers in this group. The DC will send the TC message using its LTE module to the concerned eNodeB with a specified DC reporting rate (DRR).

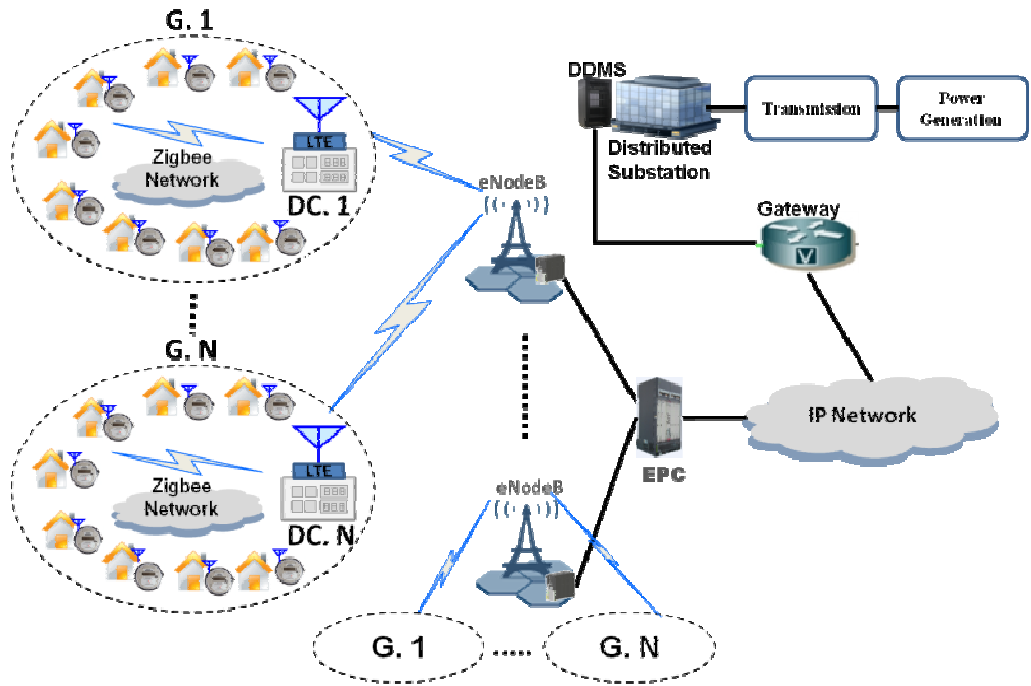

Figure 2. The suggested grouped system architecture.

In order to maintain the privacy for each customer, each DC will transmit also individually the AMI message of each SM ($SM_1$ to $SM_{K/N}$) which has his actual consumption in a scheduled manner. In the proposed system, we will schedule each DC to send SMs readings in his group with a certain reporting rate that is presented as SM reporting rate (SRR). On the other hand, DC will send the TC message every time it is polled, i.e. it sends TC message with DRR. Each time the DC transmits TC message accomplished by a set of SM readings messages according to their role in the local group schedule. Consequently, the number of transmissions required by each DC to guarantee that the control center in the corresponding distributed substation receives one reading message from each SM in the group is equal to the DRR divided by SRR. The scheduled messages send by DC is shown in Figure 3.

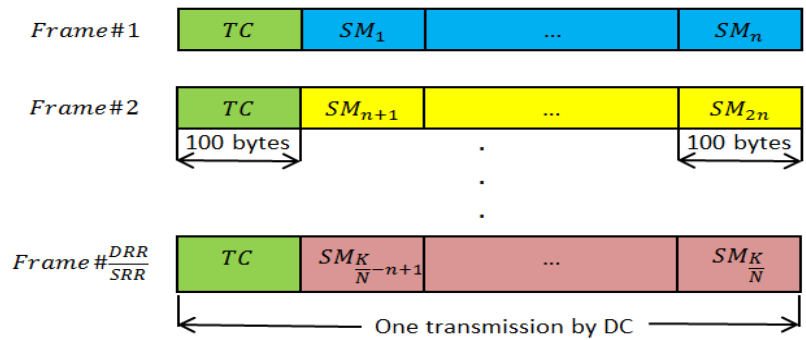

Figure 3. Scheduled messages send by DC.

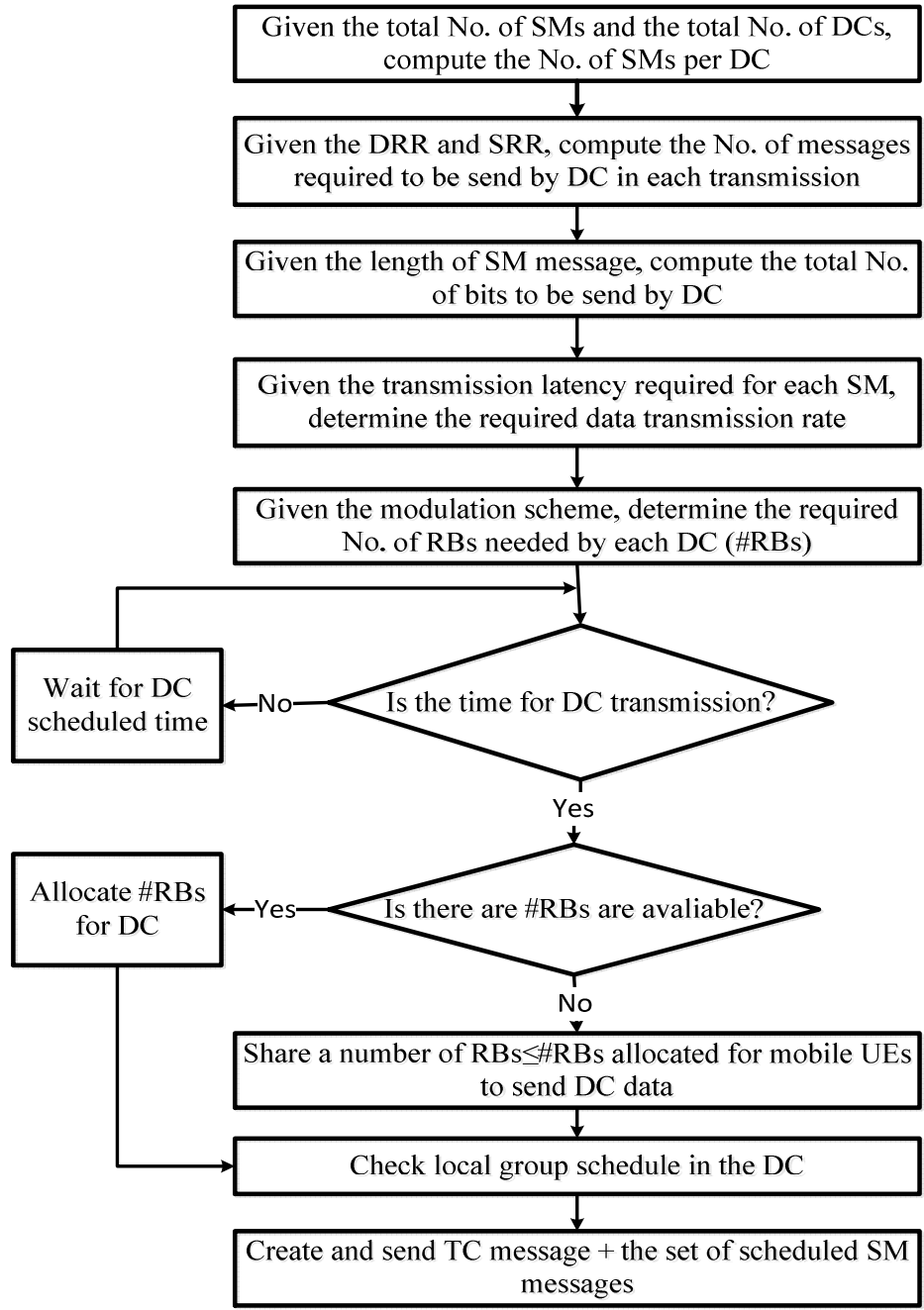

Figure 4. The proposed grouped scheduling technique.

In our solution, we propose that these scheduled messages will be send on the allocated RBs for uplink transmission. If the available number of free RBs equal to, or greater than, the required RBs to transmit the DC scheduled messages (#RBs), the eNodeB will allocate #RBs directly for them. Otherwise, if there is no enough free RBs, the eNodeB will choose a number of $RBs, \leq$ $\#RBs$, that are already assigned to mobile users to share their data with the DC data. The flowchart of the whole proposed grouped architecture and scheduling technique is showed in Figure 4.

Each eNodeB will send the received messages from the DC to the Distributed Data Management System (DDMS), which act as the control center of the distributed substation, through the LTE network. According to the total consumption collected from all sites, DDMS can take the right decision in a real time manner. The DDMS will send back its decision in form of a control signal through the LTE network to the concerned eNodeB which in turn forward this decision command to the involved DC. According to the type of the control signal, the DC determine if it must take a certain action in its area or forward the received control signal to the required SM to apply this command in that building.

Finally and as a preliminary results, the proposed architecture is practically much better than that the flat one. It dramatically reduces the required numbers of LTE modules (from $K$ to $K/N$) that used to connect the total number of SMs to the LTE network. The presented system also open the door to practically accept the idea of employing the LTE network as a communication infrastructure for smart grid application without fear of the impact of such recruitment on the performance of the LTE network. Furthermore, the suggested solution ensures a real time monitor and control for both individual customers as well as the grouped devices.

## 3. SYSTEM MODEL

The basic transmission scheme in LTE is based on Orthogonal Frequency Division Multiplexing (OFDM) in which Orthogonal Frequency Division Multiple Access (OFDMA) is used for the downlink transmission direction while Single Carrier-FDMA (SC-FDMA) is used for uplink transmission. In the time domain, LTE transmissions are organized into frames of length 10msec each. Each frame is divided into ten equally sized subframes of length 1msec which it called transmission time interval (TTI). Each subframe is divided into two equally sized slots of length 0.5msec, with each slot consisting of a number of OFDM symbols including cyclic prefix as illustrated in Figure 5.

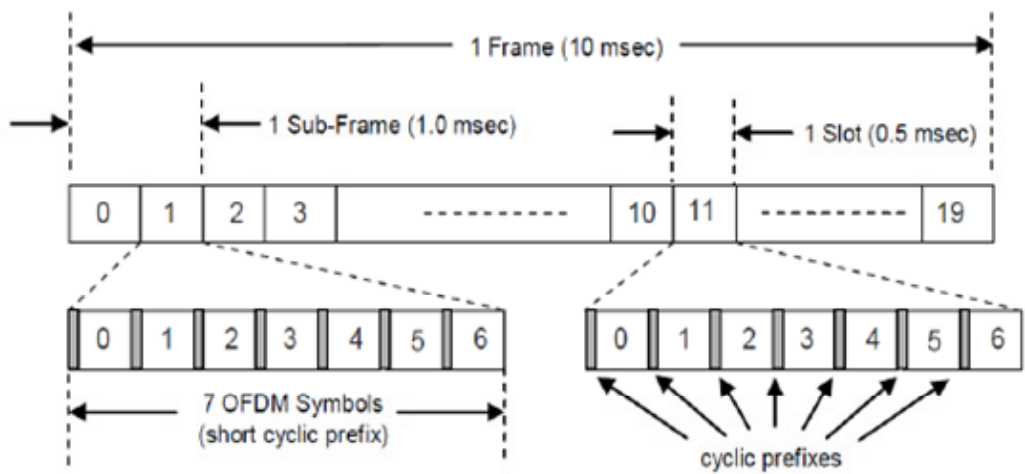

Figure 5. LTE time domain frame structure.

A resource element, consisting of one subcarrier during one OFDM symbol, is the smallest physical resource in LTE. Furthermore, as illustrated in Figure 6, resource elements are grouped into RBs, where each RB consists of 12 consecutive subcarriers in the frequency domain and one 0.5msec slot in the time domain.

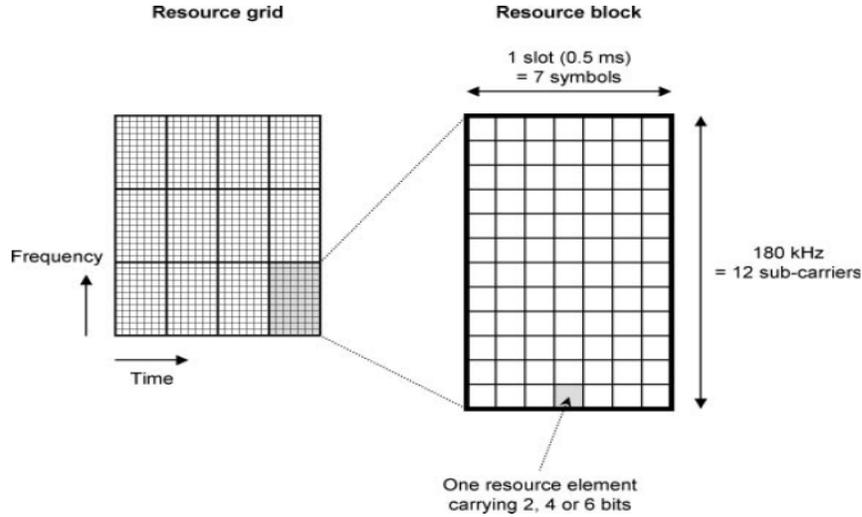

Figure 6. LTE physical time–frequency resource [23].

The number of RBs available for eNodeB is varied according to the transmission bandwidth for an eNodeB. The largest transmission bandwidth for an eNodeB is 20 MHz, which provides 100 RBs while the smallest transmission bandwidth for is 1.4 MHz, which provides 6 RBs [24].

In the following subsections, an analytical model is introduced to calculate the amount of bandwidth consumed in uplink transmission for both flat and proposed grouped system architectures.

### 3.1 Flat Architecture

Let, $K_{SM}^{Total}$ indicates the total number of SMs in a given area. In flat architecture, each SM is assumed to be enabled by LTE interface and configured to transmit its AMI reading data directly to the eNodeB every $T^{SM}$ second who forwards it to the control center. Let $L^{SM}$ represents the length of AMI reading message send by each SM indicating the amount of energy consumption for appliances connected to it. If the number of TTIs required by each SM to transmit their data to eNodeB is $N_{TTI}^{SM}$. So, the number of RBs that are allocated for each SM in uplink transmission $N_{RB,UL}^{SM}$ should be selected to satisfy the following inequality.

$$R\left(I_{MCS}, N_{RB,UL}^{SM}\right) \geq \frac{L^{SM}}{N_{TTI}^{SM} * TTI * N_{TS}^{TTI}} \qquad (1)$$

Where $R\left(I_{MCS}, N_{RB,UL}^{SM}\right)$ is the data rates (bps) carried by the $N_{RB,UL}^{SM}$ resource blocks which it depends on the chosen modulation code scheme index $I_{MCS}$ and $N_{TS}^{TTI}$ indicates the number of time slots per TTI. Given the system configuration parameters in terms of AMI reading message length and the required number of TTI, the minimum number of RBs needed by each SM to transmit their data to eNodeB can be determined from the transport block size table [25] to achieve the inequality in (1).

In order to evaluate effect of SMs AMI traffic on the cellular User Equipment's (UEs), the amount of bandwidth consumed by the total number of SMs to send their AMI reading messages relative to the total available bandwidth by the LTE eNodeB is calculated. Let TB indicates the transmission bandwidth used by eNodeB and $N_{RB}^{TB}$ is the number of RBs available per each transmission bandwidth.

Let $\eta^{UL} = (N_{SF}^{UL}/N_{SF}^{Frame})$ represents the percentage of bandwidth allocated for uplink transmission in LTE-TDD network where, $N_{SF}^{UL}$ indicates the total number of subframes allocated for uplink transmission in each LTE frame, and $N_{SF}^{Frame}$ represents the total number of subframes per each LTE frame. The value of $\eta^{UL}$ can be determined according to the chosen uplink-downlink configuration set in LTE-TDD network. Define the total number of smart meters that will share the same resource blocks during one $T^{SM}$ as $N_{SM}^{RB} = \frac{T^{SM}}{T_{Service}}$, where $T_{Service}$ is the maximum latency required to transmit the smart meter readings. Therefore, the amount of bandwidth consumed by the total number of SMs in flat architecture relative to the total available bandwidth by eNodeB for uplink transmission is represented by $\Gamma^{Flat}$ and can be calculated as follows.

$$\Gamma^{Flat} = \frac{\left(K_{SM}^{Total}\right)^2 * N_{RB,UL}^{SM} * (T_{Service})^2}{N_{RB}^{TB} * (T^{SM})^2 * \eta^{UL}} \tag{2}$$

Where $N_{RB}^{TB}$ can be determined according to LTE standard [24] using (3).

$$N_{RB}^{TB} = \begin{cases} 6, & If\ TB = 1.4\ MHz \\ 15, & If\ TB = 3\ MHz \\ 25, & If\ TB = 5\ MHz \\ 50, & If\ TB = 10\ MHz \\ 75, & If\ TB = 15\ MHz \\ 100, & If\ TB = 20\ MHz \end{cases} \tag{3}$$

## 3.2 Grouped Architecture

In grouped architecture, the readings for each group of SMs are transmitted and collected by a specific DC. Each DC is assumed to be enabled by LTE interface and configured to transmit its AMI reading data directly to the eNodeB every $T^{DC} = (1/DRR)$ seconds who forwards it to the control center. The DC and its associated group of SMs are assumed to be connected through short distance communication technology like Zigbee. The AMI data readings collected by each DC from its group of SMs are used to calculate the total energy consumption related to that group of meters.

Let $K_{DC}^{Total}$ represents the total number of DCs. Then the number of SMs served by each data concentrator $K_{SM}^{DC}$ can be calculated as $K_{SM}^{DC} = K_{SM}^{Total}/K_{DC}^{Total}$. In each time the DC is polled, a set of AMI messages are sent to the control center. One of these set is an AMI message with length of $L^{SM}$ bits indicating the total energy consumption for the group of meters served by that DC. The remaining set are AMI reading messages for a selected number of SMs from the group. This number of SMs is determined in such a way that the control center receives one reading message from each SM in the group in a sequential scheduled manner every $\tilde{T}^{SM} == (1/SRR)$ seconds. Therefore, the number of SMs per DC needs to send their AMI reading messages every time the DC is polled can be calculated using (4)

$$\tilde{K}_{SM}^{DC} = \frac{K_{SM}^{DC} * T^{DC}}{\tilde{T}^{SM}} \tag{4}$$

Therefore, the amount of data bits $L^{DC}$ needs to be transmitted by each DC every time is polled can be written as follows.

$$L^{DC} = L^{SM} \left(1 + \frac{K_{SM}^{DC} * T^{DC}}{\check{T}^{SM}}\right) \tag{5}$$

$$L^{DC} = \frac{L^{SM} * K_{DC}^{Total} * \check{T}^{SM} + L^{SM} * K_{SM}^{Total} * T^{DC}}{K_{DC}^{Total} * \check{T}^{SM}} \tag{6}$$

Let $N_{TTI}^{DC}$ represents the number of TTIs required for each DC to transmit their data to eNodeB. So, the number of RBs that allocated for each DC in uplink transmission $N_{RB}^{DC}$ should be selected to satisfy the following inequality.

$$R\left(I_{MCS}, N_{RB,UL}^{DC}\right) \geq \frac{L^{DC}}{N_{TTI}^{DC} * TTI * N_{TS}^{TTI}} \tag{7}$$

$$R\left(I_{MCS}, N_{RB,UL}^{DC}\right) \geq \frac{L^{SM} * K_{DC}^{Total} * \check{T}^{SM} + L^{SM} * K_{SM}^{Total} * T^{DC}}{K_{DC}^{Total} * \check{T}^{SM} * N_{TTI}^{DC} * TTI * N_{TS}^{TTI}} \tag{8}$$

Where $R\left(I_{MCS}, N_{RB,UL}^{DC}\right)$ is the data rates (bps) carried by the $N_{RB,UL}^{DC}$ resource blocks. Given the total number of SMs, the total number of DCs, and the system configuration parameters, the minimum number of RBs needed by each smart meter to transmit their data to eNodeB can be determined from the transport block size table [23] to achieve the inequality in (8).

Assume that the total number of data concentrators that will share the same resource blocks during one $T^{DC}$ is $N_{DC}^{RB} = \frac{T^{DC}}{T_{Service}}$ where in this case $T_{Service}$ is the maximum latency required to transmit the DC data. Hence, the amount of bandwidth consumed by the total number of SMs to send their AMI reading messages in grouped architecture relative to the total available bandwidth by eNodeB is represented by $\Gamma^{Grouped}$ and can be calculated as follows.

$$\Gamma^{Grouped} = \frac{\left(K_{DC}^{Total}\right)^2 * N_{RB,UL}^{DC} * (T_{Service})^2}{N_{RB}^{TB} * (T^{DC})^2 * \eta^{UL}} \tag{9}$$

In order to evaluate the enhancement in the consumed bandwidth reduction due to the grouped architecture; $BRR$ is defined as the bandwidth reduction ratio and calculated as follows.

$$BRR = \frac{\Gamma^{Grouped}}{\Gamma^{Flat}} = \frac{\left(K_{DC}^{Total}\right)^2 * N_{RB,UL}^{DC} * (T^{SM})^2}{\left(K_{SM}^{Total}\right)^2 * N_{RB,UL}^{SM} * (T^{DC})^2} \tag{10}$$

$$BRR = \frac{N_{RB,UL}^{DC} * (T^{SM})^2}{(K_{SM}^{DC})^2 * N_{RB,UL}^{SM} * (T^{DC})^2} \tag{11}$$

In order to make the comparison fair, the number of TTIs required by each SM and each DC are set with equal values ($N_{TTI}^{SM} = N_{TTI}^{DC}$) as well as the periodic polling time for each SM and each DC ($T^{SM} = T^{DC}$). In this case, (11) is reduced as follows.

$$BRR = \frac{N_{RB,UL}^{DC}}{(K_{SM}^{DC})^2 * N_{RB,UL}^{SM}} \qquad (12)$$

It is obvious from (12) that the number of SMs per DC is a crucial parameter with high influence on the performance of the grouped architecture. It is also important to note that $K_{SM}^{DC}$ and $N_{RB,UL}^{DC}$ are dependent variables as the change in the number of SMs per DC will lead to change on the amount of data need to be send by each DC and so different number of RBs are required by each DC.

## 4. RESULTS AND DISCUSSION

In this section, the analytical model presented in section III is used to evaluate the performance of the proposed system. The percentage of bandwidth consumed that needed to send the AMI reading data for all SMs to eNodeB is evaluated and compared in both flat and grouped system architectures. Also the effect of different system parameters such as the total number of SMs need to be served and the number of SMs per DC on the system performance is investigated.

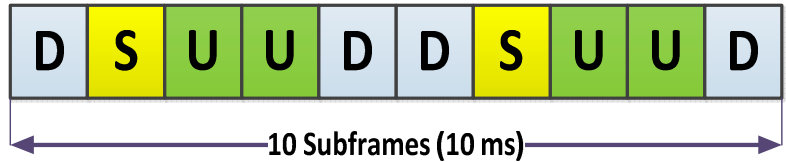

Figure 7. Uplink-downlink Configuration.

Table 1. System configuration parameters and assumptions.

| Parameter | Value | Parameter | Value |
|---|---|---|---|
| $L^{SM}$ | 100 byte | TB | 1.4, 3, 5, 10, 15, and 20 MHz |
| $T^{SM}$ | 15 minutes | $N_{RB}^{TB}$ | 6, 15, 25, 50, 75, and 100 |
| $T^{DC}$ | 15 minutes | $N_{SF}^{Frame}$ | 10 |
| $\tilde{T}^{SM}$ | 60 minutes | $N_{SF}^{UL}$ | 4 |
| $T_{Service}$ | 1 Second | $I_{MCS}$ | QPSK, 16QAM, and 64QAM |
| $N_{TTI}^{SM}$ | 8 | | |
| $N_{TTI}^{DC}$ | 8 | $K_{SM}^{Total}$ | Variable |
| $N_{TS}^{TTI}$ | 2 | $K_{DC}^{Total}$ | Variable |
| TTI | 1 msec | $K_{SM}^{DC}$ | Variable |

We assume that time division duplexing (TDD) is used in LTE network with uplink-downlink configuration on eNodeB is set as 1; that is 4 TTIs for uplink (U), 4 TTIs for downlink (D) subframe, and 2 TTIs for special subframe (S) as shown in Figure.7. Quadrature phase shift keying (QPSK) modulation scheme is chosen for SM data transmission since it has the lowest bit error rate (BER) compared with 16 QAM, or 64 QAM modulation schemes especially that SMs do not need to send high-speed data. Therefore, the modulation code scheme index $I_{MCS} = 9$ is selected. Table 1 summarizes the system configuration parameters and assumptions.

Let the coverage area for one eNodeB is 10 square kilometer. As the number of SMs in a given service area is depends on the nature of this area (urban, suburban, and rural) and the density of buildings on it as well as the nature of buildings (single flat or multi-flat). So, we assumed that the number of SMs in one square kilometer ranges from 800 to 18000 SMs. This allows us to evaluate the performance of the proposed system in both low and high density service areas.

Figure 8, shows the percentage of bandwidth consumed in both flat and grouped architecture for different number of SMs. As can be seen, in the flat scenario the consumed bandwidth is increased dramatically as the number of SMs increased to the extent that 4500 SMs consume all the available bandwidth of the LTE which means if this is acceptable no more than this value can be served by the flat architecture. On the other hand, the proposed system using grouped architecture outperforms the flat architecture and consumes less amount of bandwidth. While we can get a percentage of only 0.3 % from the total available bandwidth for uplink transmission to serve 4500 SMs using 30 SMs/DC or more, the maximum BW consumption percentage does not exceed 2 % for 10 SMs/DC. The previous case (up to 4500 SMs) can be considered as an example for the low density areas meaning that the flat architecture is not applicable for high density areas.

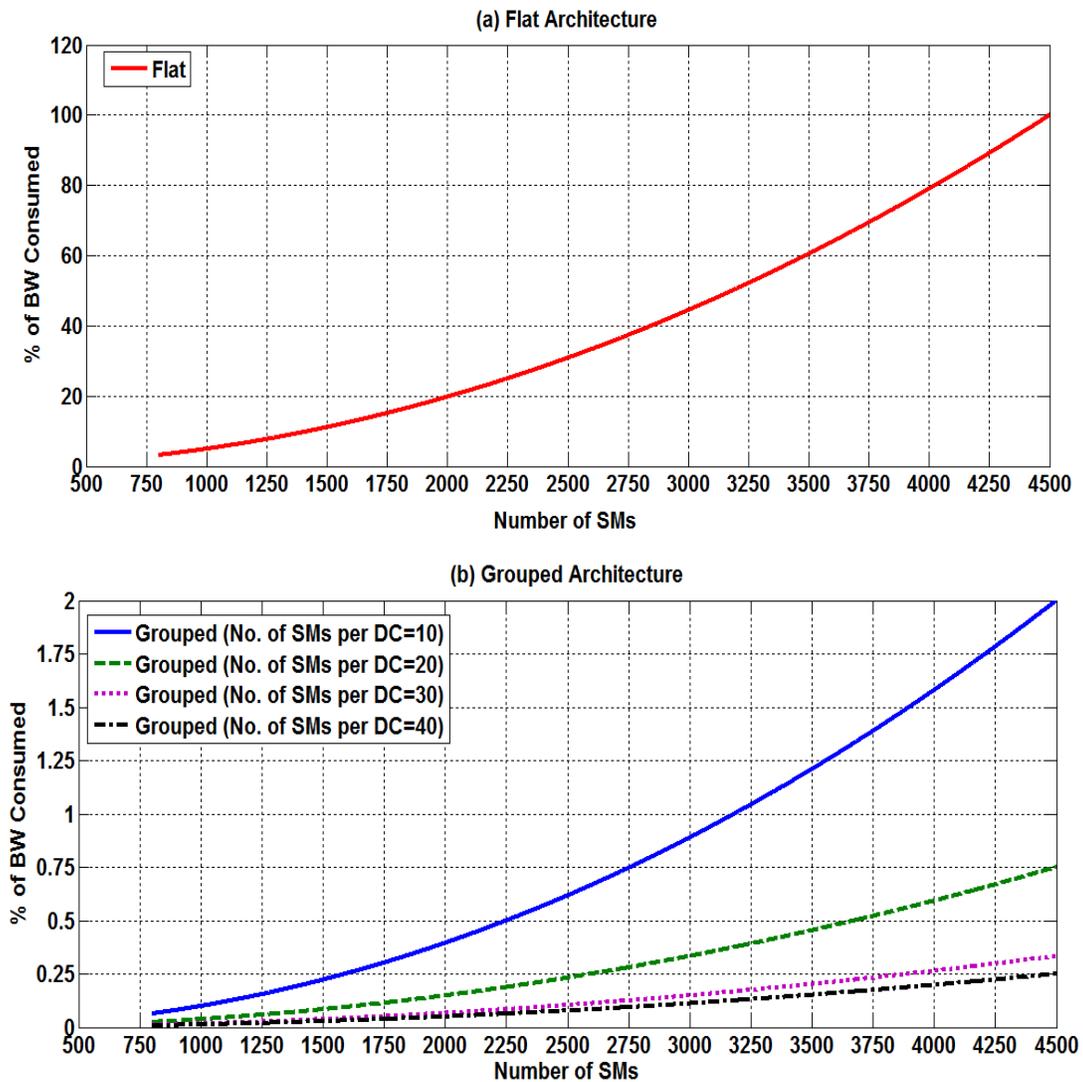

Figure 8. Percentage of bandwidth consumed at different number of SMs (Low density case).

In contrast, in the high density areas the proposed grouped architecture will consumes less than 6 % for 40 SMs/DC and 33 % in the worst case of 10 SMs/DC to serve a total of 18000 SMs as shown in Figure 9.

It is also important to note that the number of SMs per DC is a significant parameter that effect on the performance of the proposed system especially as the density of SMs increased. This results also shows that in the worst when the number of SMs per DC equal 10, the percentage of the bandwidth consumed is ranges from 0.0035% in case of low density (800 SMs) to 0.16% in case of high density (18000 SMs) which are very small percentages. This shows that the presented grouped system can able to be used in areas having high density of SMs with little effect on the required QoS levels for various traffic types of mobile users in the LTE network.

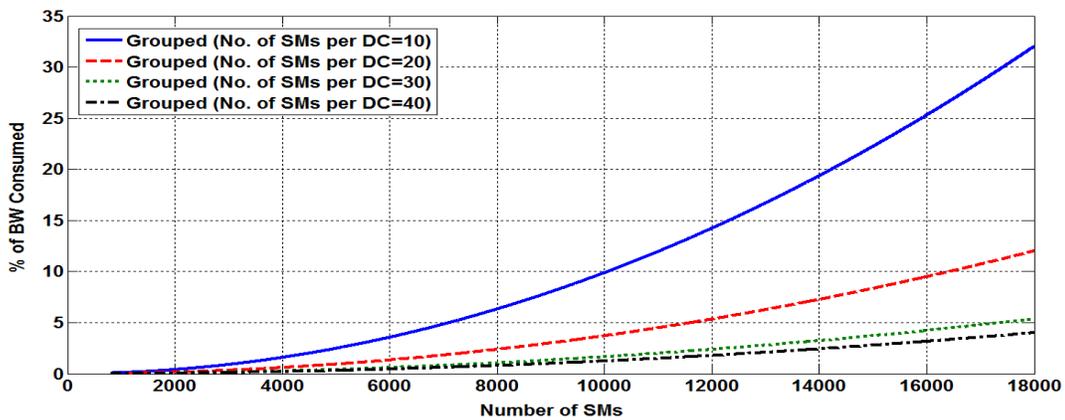

Figure 9. Percentage of bandwidth consumed at different number of SMs (High density case).

As a figure of merit to investigate the enhancement that can be achieved by using the grouped architecture, we will compute the Bandwidth Reduction Ratio (BRR) as the ratio between the consumed bandwidth in grouped architecture and in flat one. Figure 10, shows the effect of the number of SMs per DC on BRR. As we can see BRR is highly improved as the number of SMs per DC increased from 4 to 40 where BRR is dropped greatly from 0.062 to 0.012. This can be explained by two reasons; the first one is that the increase in the number of SMs per DC will lead to reduction in the number of required DCs. The other reason is that the increase ratio in the required bandwidth per DC due to the increase in the number of SMs per DC is smaller than the reduction ratio in the required number of DCs. This because a large number of SMs readings data are grouped together and sent in one AMI reading message with the total energy consumption.

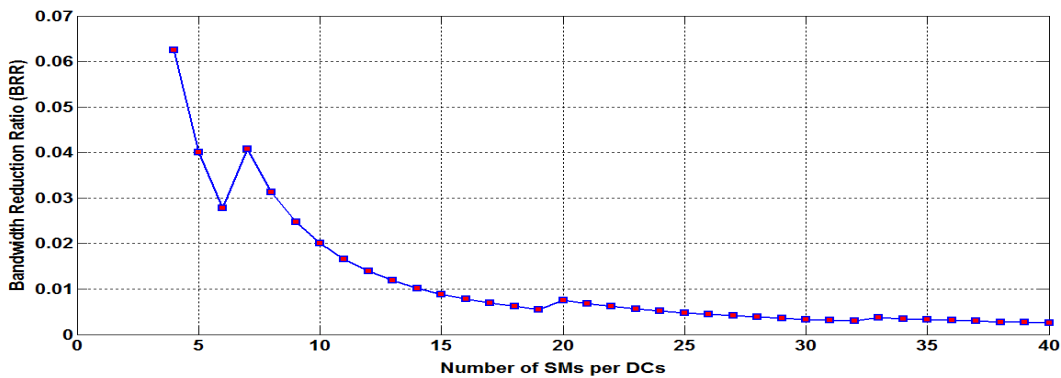

Figure 10. BRR Vs. number of SMs per DC.

It is also important to note that for a given number of SMs per DC, the BBR is not affected by changing the total number of SMs. This because the increase in the total number of SMs corresponding to flat architecture is compensated by the increase in the total number of required DCs corresponding to the grouped architecture. In addition, we can conclude that there is no benefit to increase number of SMs/DC than 20 because the improvement in BRR after this value is minimal and actually we have already achieved what is required.

The previous results lead us to study the effect of the total number of DCs on the performance of the proposed system where for a fixed number of SMs/DC (e.g. 20) the number of DCs will be increased as the number of SMs increased. The relation between BBR and the total number of SMs for a given value of the total number of DCs can be plotted as shown in Figure.11. As seen, the BRR reduces as the total number of SMs increases for a fixed number of DCs. This means that the amount of bandwidth consumed by the grouped architecture relative to the bandwidth consumed by the flat architecture is decreased as the number of SMs increased. This indicates that the proposed grouped architecture system works more efficiently in service areas having high density of SMs compared to the flat architecture system. It is also shown that the BBR is decreased as the total number of DCs decreased for a given number of SMs. This is because the number of SMs served by each DC is increased and so a larger number of AMI reading data can be grouped together in one reading message which consumes less bandwidth compared to the flat architecture case.

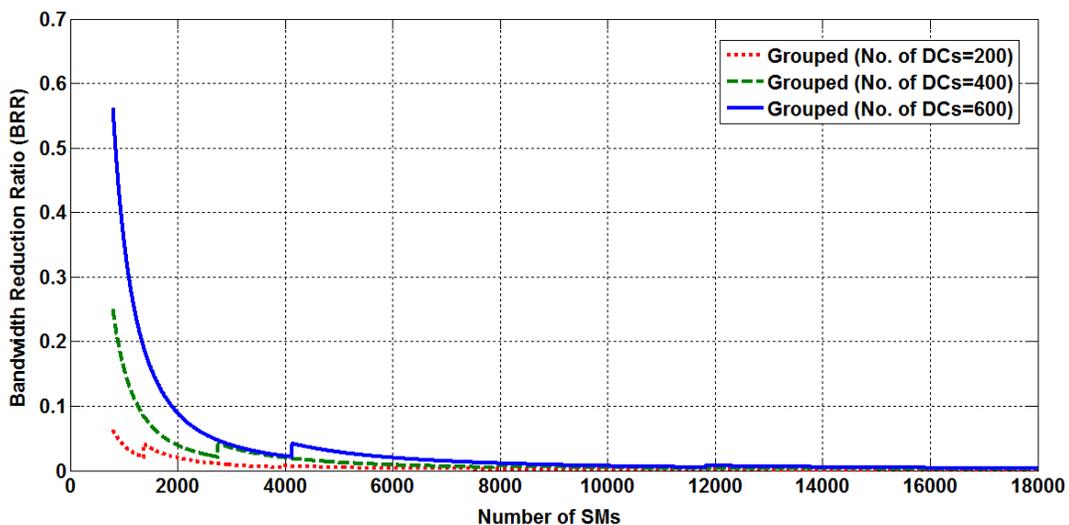

Figure 11. BRR Vs. number of SMs for different number of DCs.

Tables 2 and 3 show the percentage of bandwidth consumed at different system bandwidth using various modulation techniques in both low and high density cases. For the lowest available bandwidth of one eNodeB in LTE (1.4 MHz), the maximum number of SMs that can be served is 1550 which will almost consume the total bandwidth which is considered a low density case. In this case, the consumption in flat architecture is almost independent on the modulation technique whatever the system bandwidth. This is because a minimum number of RBs (One RB) using QPSK modulation is quietly sufficient to transmit the smart meter reading message and using higher order modulation techniques can't provide any reduction on the required number of RBs. On the other hand, the modulation technique has significant effect on the bandwidth consumption in case of grouped architecture where the bandwidth consumption of 64 QAM is almost quarter of the value of 16 QAM.

For high density case (18000 SMs), the bandwidth consumption of flat architecture is overestimated even for the highest system bandwidth offered by LTE. In contrast, for the worst channel condition where QPSK modulation is used, the grouped architecture almost consumes one third of the system bandwidth for 1.4 MHz and only 2% for 20 MHz. Finally and as the best case we can reach the percentage of 0.5% for 64 QAM and 20 MHz.

Table 2. Percentage of bandwidth consumed (low density case: No. of SMs=1550).

| System BW | QPSK | | | | 16 QAM | | | | 64 QAM | | | |
|---|---|---|---|---|---|---|---|---|---|---|---|---|
| | Flat | SMs/DC=10 | SMs/DC=20 | SMs/DC=40 | Flat | SMs/DC=10 | SMs/DC=20 | SMs/DC=40 | Flat | SMs/DC=10 | SMs/DC=20 | SMs/DC=40 |
| 1.4 MHz (6 RBs) | 98.868 | 1.977 | 0.742 | 0.247 | 98.868 | 0.989 | 0.494 | 0.124 | 98.868 | 0.989 | 0.247 | 0.062 |
| 3 MHz (15 RBs) | 39.547 | 0.791 | 0.297 | 0.099 | 39.545 | 0.396 | 0.198 | 0.049 | 39.547 | 0.396 | 0.010 | 0.025 |
| 5 MHz (25 RBs) | 23.728 | 0.475 | 0.178 | 0.059 | 23.728 | 0.237 | 0.119 | 0.030 | 23.728 | 0.237 | 0.059 | 0.015 |
| 10 MHz (50 RBs) | 11.864 | 0.237 | 0.089 | 0.030 | 11.864 | 0.119 | 0.059 | 0.015 | 11.864 | 0.119 | 0.030 | 0.007 |
| 15 MHz (75 RBs) | 7.910 | 0.158 | 0.059 | 0.020 | 7.910 | 0.0791 | 0.040 | 0.010 | 7.910 | 0.079 | 0.020 | 0.005 |
| 20 MHz (100 RBs) | 5.932 | 0.119 | 0.045 | 0.015 | 5.932 | 0.059 | 0.030 | 0.007 | 5.932 | 0.059 | 0.015 | 0.004 |

Table 3. Percentage of bandwidth consumed (High density case: No. of SMs=18000).

| System BW | QPSK | | | | 16 QAM | | | | 64 QAM | | | |
|---|---|---|---|---|---|---|---|---|---|---|---|---|
| | Flat | SMs/DC=10 | SMs/DC=20 | SMs/DC=40 | Flat | SMs/DC=10 | SMs/DC=20 | SMs/DC=40 | Flat | SMs/DC=10 | SMs/DC=20 | SMs/DC=40 |
| 1.4 MHz (6 RBs) | - | - | 100.00 | 33.33 | - | - | 66.67 | 16.67 | - | - | 33.33 | 8.33 |
| 3 MHz (15 RBs) | - | - | 40.00 | 13.33 | - | 53.33 | 26.67 | 6.67 | - | 53.33 | 13.33 | 3.33 |
| 5 MHz (25 RBs) | - | 64.00 | 24.00 | 8.00 | - | 32.00 | 16.00 | 4.00 | - | 32.00 | 8.00 | 2.00 |
| 10 MHz (50 RBs) | - | 32.00 | 12.00 | 4.00 | - | 16.00 | 8.00 | 2.00 | - | 16.00 | 4.00 | 1.00 |
| 15 MHz (75 RBs) | - | 21.33 | 8.00 | 2.67 | - | 10.67 | 5.33 | 1.33 | - | 10.67 | 2.67 | 0.67 |
| 20 MHz (100 RBs) | - | 16.00 | 6.00 | 2.00 | - | 8.00 | 4.00 | 1.00 | - | 8.00 | 2.00 | 0.50 |

## 5. CONCLUSION

Smart grids, created to optimize the energy efficiency of power grids, have a number of challenges. One of these challenges is the requirement of high efficient and reliable communication networks to achieve the real time monitoring and control on both generation and customer sides. 4G LTE, a standard by 3GPP, is capable of supporting mobile user services with high data rates as well as high speed mobility. Resounding and rapid success of LTE leads to an increased concern to use it for different applications domain including smart grids. In this paper we proposed a grouped hierarchal architecture for smart grid communications in LTE network. We also presented an efficient grouped scheduling technique for SMs communications over LTE infrastructure. The results we have obtained emphasize that our suggested system increases the opportunities for using LTE network as the communication infrastructure for smart grid traffic. The results show that the bandwidth consumed by the proposed grouped system is almost 33 % from that required by the flat architecture in low density areas. This value reduced to 20% in case of high density areas, which lead to using our solution in areas having high density of SMs with little effect on the required QoS for various traffic types in the LTE network.

As a future work, we are currently investigating how the RBs required for SG traffic can be dynamically allocated in a shared manner with the assigned RBs for mobile services that has minimum required QoS.